\documentclass{emulateapj}
\usepackage{txfonts}
\usepackage{graphicx}
\usepackage{nicefrac}
\usepackage{float}
\usepackage{color}

\shorttitle{Size of the Accretion Disk in the Gravitationally Lensed Quasar SDSS 
J1004+4112}
\shortauthors{Fian et al.}
\begin{document}

\title{Size of the Accretion Disk in the Gravitationally Lensed Quasar SDSS 
J1004+4112 from the Statistics of Microlensing Magnifications}

%%   \date{Received December 17, 2014; accepted .......}

\author{C. Fian\altaffilmark{1,2}, E. Mediavilla\altaffilmark{1,2}, A. Hanslmeier\altaffilmark{3}, A. Oscoz\altaffilmark{1}, M. Serra-Ricart\altaffilmark{1,2}, J. A. Mu\~{n}oz\altaffilmark{4,5}, J. Jim{\'e}nez-Vicente\altaffilmark{6,7}}

\altaffiltext{1}{Instituto de Astrof\'{i}sica de Canarias (IAC), V\'{i}a L\'{a}ctea S/N, La Laguna E 38200, Tenerife, Spain}
\altaffiltext{2}{Departamento de Astrof\'{i}sica, Universidad de La Laguna, E-38200 La Laguna, Spain}
\altaffiltext{3}{Institute of Physics (IGAM), University of Graz, Universit{\"a}tsplatz 5, 8010, Graz, Austria}
\altaffiltext{4}{Departamento de Astronom\'{i}a y Astrof\'{i}sica, Universidad de Valencia, E-46100 Burjassot, Valencia, Spain}
\altaffiltext{5}{Observatorio Astron\'{o}mico, Universidad de Valencia, E-46980 Paterna, Valencia, Spain} 
\altaffiltext{6}{Departamento de F\'{i}sica Te\'{o}rica y del Cosmos. Universidad de Granada, Av. Fuentenueva s/n, E-18071, Granada, Spain}
\altaffiltext{7}{Instituto Carlos I de F\'{i}sica Te\'{o}rica y Computacional, Universidad de Granada, Av. Fuentenueva s/n, E-18071, Granada, Spain}

\begin{abstract}
We present eight monitoring seasons of the four brightest images of the gravitational lens SDSS J1004+4112 observed between December 2003 and October 2010. Using measured time delays for the images A, B and C and the model predicted time delay for image D we have removed the intrinsic quasar variability, finding microlensing events of about 0.5 and 0.7 mag of amplitude in the images C and D. From the statistics of microlensing amplitudes in images A, C, and D, we have inferred the half-light radius (at $\lambda_{rest}=2407\AA$) for the accretion disk using two different methods, $R_{1/2}=8.7^{+18.5}_{-5.5} \sqrt{M/0.3 M_\odot}$ (histograms product) and $R_{1/2} = 4.2^{+3.2}_{-2.2} \sqrt{M/0.3 M_\odot}$ light-days ($\chi^2$). The results are in agreement within uncertainties with the size predicted from the black hole mass in SDSS J1004+4112 using the thin disk theory.
\end{abstract}


\section{Introduction}
The flux variability of the images of a gravitationally lensed quasar is a combination of the intrinsic variability of the source, correlated by a time delay between the different images, and gravitational microlensing that depends on the random distribution of stars in the lens galaxy. This last variability is 
uncorrelated between images (\citealt{Chang1979}; see also the review by \citealt{wam2006}). The analysis of lensed quasars light curves has important applications in cosmology (determination of time delays to infer the Hubble constant, \citealt{Ref1964}) and in the study of quasars structure (\citealt{Chang1979,Chang1984}; see also \citealt{Kochanek2004} and \citealt{wam2006}). In this paper we will focus in the last application, using microlensing statistics to determine the quasar accretion disk size \citep{Pool2007,fohl2007,fohl2008a,fohl2008b,Morgan2010,Sluse2011,Blackburne2011,Motta2012,Blackburne2014,Blackburne2015,Mosquera2011,Jimenez2012,Jimenez2014,Hainline2013,Mosquera2009,Mosquera2013,MacLeod2015,Jimenez2015b,Jimenez2015a,Munoz2015,Mediavilla2015}.\\

The wide-separation lensed quasar SDSS J1004+4112 is one of the rare examples of a quasar lensed by an intervening galaxy cluster \citep{wam2003,inada2006}. It is lensed into five images \citep{inada2005}, with a maximum image separation of 14.62$''$. The quasar has a redshift of $z_s= 1.734$ and the redshift of the galaxy cluster is $z_
l= 0.68$ \citep{fohl2008a}. The lag between images A, B, and C has been experimentally measured by \citet{fohl2008a} who obtained a time delay of 40.6 days between components A and B and of 822 days between A and C (the largest delay measured for a gravitational lens system). It has not been yet possible to measure the time delay of image D. From a model of SDSS J1004+4112 in which the mass distribution of the system is revisited, \citet{oguri2010} predicts a time delay between A and D of 1218 days.\\

After correcting for the time delays and mean magnitude differences, \citet{fohl2008a} found clear indications of microlensing flux variability in the residuals of the light curves ( i.e. the differences between the observed light curves and the modeled intrinsic variability of the quasar). They fitted the residual light curves of image A using a model of microlensing flux variability based in magnification maps to estimate the accretion disk size of the lensed quasar, obtaining a small size below the predictions of thin disk theory.\\

In this paper we are presenting seven years, corresponding to eight observational seasons, of optical monitoring data for the four brightest images of SDSS J1004+4112 spanning 2505 days from December 2003 to October 2010, partly coincident with the epochs studied by \citet{fohl2008a}, but that significantly extend the coverage with around 1200 additional days. Our aim is to use these data to study the existence of possible microlensing events and to estimate the size of the accretion disk of the lensed quasar. We have been able to make microlensing analysis including image D for the first time thanks to the availability of long enough observations to correct for the time delay shifting the light curves.\\

Instead of the light curve fitting method (see e.g. \citealt{Kochanek2004}) we will follow the single epoch method combined with the flux ratios of a large enough source in the quasar as to be insensitive to microlensing to establish the baseline for no microlensing magnification (see, e.g., \citealt{Mediavilla2009}). Based on single epoch spectroscopy \citet{Motta2012} and \citet{Jimenez2014} obtained disk size estimates larger than the results by \citet{fohl2008a}. In the present work we will extend the single epoch method to all the epochs in the available light curves, increasing the statistical significance. The basic idea is to compare the histogram of microlensing magnification obtained from the observations corresponding to a time interval with the simulated predictions of microlensing variability for sources of different sizes. This comparison will allow to evaluate the likelihood of the different values adopted for the size. We will use the optical light curves, with coverage extended by us, to infer microlensing flux variability and the midIR data from \citet{Ross2009} to determine the baseline for no microlensing variability.\\

The paper is organized as follows. In $\S2$ we present the data and in $\S3$ the light curves of each image. The estimate of the quasar accretion disk size based on the statistics of microlensing magnifications is discussed in $\S4$. Finally, the main results are summarized in $\S5$.

\section{Data and Observations}
The photometric monitoring presented in this paper took place between December 2003 and October 2010. We monitored SDSS J1004+4112 in the Johnson-Bessell's R-band using the 82 cm telescope (IAC80) at the Instituto de Astrof\'{i}sica de Canarias’ Teide Observatory (Tenerife, Canary Islands, Spain). Two different CCD were used. From 2003 to 2005 a Thomson 1024 x 1024 chip was employed, giving a field of view of about 7$\farcm$5, with a pixel size of 0 $\farcs$43. Since 2005 a new CCD, CAMELOT, was installed. CAMELOT hosts a E2V 2048 x 2048 chip with 0 $\farcs$304 pixels, corresponding to a 10.4 x 10.4 arcmin$^2$ FOV. A standard R broadband filter was always used for the observations, fairly close to the Landolt R \citep{Landolt1992}. Due to the instrumental change a constant gap of $\sim$ 0.2 mag appeared in the light curves that has been removed taking the new instrumental system as reference. The combined data set consists of 109 epochs (i.e. 109 nights) and the average sampling rate is once every 23rd day. This large average sampling rate arises due to the relatively large seasonal gaps. There are seven seasonal gaps, the largest one stretching over 11 months from June 2007 to May 2008, i.e. a time period of more than 300 days. The mean observational cadence of the four images is $\sim$ 9 days for the first two seasons, $\sim$ 10 days for the third and fourth season, $\sim$ 6 days for the fifth season and $\sim$ 18 days for the last two seasons.\\

\begin{figure}
\centering
\includegraphics[width=4cm]{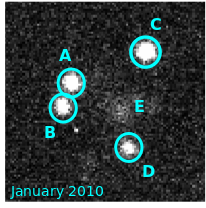}
\caption{Image of the four brightest quasar images A, B, C and D of SDSS J1004+4112. The faint spot in the middle of the images is a bright galaxy belonging to the intervening lensing cluster.\\
}
\label{snapshots}
\end{figure}

Fig. \ref{snapshots} shows an image taken with the IAC80 of the four brightest quasar images A, B, C and D of SDSS J1004+4112.
Photometric data was obtained applying a completely automatic IRAF task, {\it pho2com}, developed by \citealt{Serra1999}. This code yields accurate photometry by simultaneously fitting a stellar two-dimensional profile to each QSO component by means of DAOPHOT software. To remove inconsistent data due to instrumental problems or other sources of error related to the data reduction we have, in first place, removed the observations in which a sudden change in magnitude simultanously appears in all the images. In a second step, the magnitudes between two consecutive points were compared and if the difference in magnitude was greater than twice the standard deviation, the point was discarded. Only around 10 data points out of 109 had to be rejected and around 100 data points are left for analysis.\\

\section{Light Curves and Magnitude Changes}
\begin{figure*}
\centering
\includegraphics[width=11.6cm]{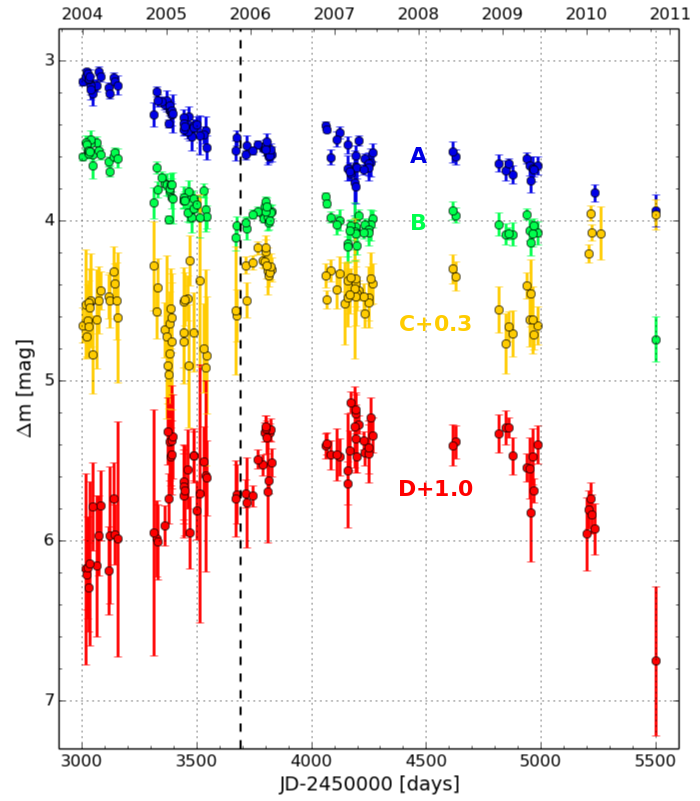}
\caption{Light curves of the four images A, B, C and D of the quasar SDSS J1004+4112 from December 2003 to October 2010. Horizontal axes show the Julian (bottom) and Gregorian (upper) dates. The vertical axis is shifted to match the photometry by \citet{fohl2008a}. The light curves of the images C and D are shifted by 0.3 mag and 1.0 mag respectively so that they do not overlap with each other. We smooth the data with a square filter of $\pm$ 5 days to reduce noise and emphasize trends.\\
}
\label{Lens1004R_lightcurves}
\end{figure*}

In Fig. \ref{Lens1004R_lightcurves} we show the resulting light curves of the quasar images A-D split over eight observing seasons. The dashed vertical line shown at 3700 days represents the change of the old CCD to the new CCD and one can clearly see that the error bars of the observations with the new CCD are smaller by almost a factor of two. The images C (yellow colored light curve) and D (red colored light curve) are shifted by 0.3 mag and 1.0 mag respectively so that they do not overlap with each other.\\

Due to our larger uncertainties, we have not been able to improve the time delay measurements of \citet{fohl2008a}, therefore we use their time delay estimations (for images A, B and C) to shift the light curves and remove intrinsic variability. For calculating the time delay between image A and B they used the polynomial fitting method and the analysis of their data yields a time delay of $\Delta t_{BA} = 40.6 \pm 1.8$ days. They also measured the time delay for the wide separated (14.62$''$) image C relative to the close image pair A/B. Using the dispersion spectra method they found $\Delta t_{CA} = 822 \pm 7$ days and $\Delta t_{CB} = 780 \pm 6$ days. D should lag the other three images but up to now no features can be seen in the light curve of image D than can be matched to the first season of image A. They derived a lower limit on the time delay between images A and D of $\Delta t_{DA} > 1250$ days (3.4 years) \citep{fohl2008a,fohl2008b}. We will use the model-predicted AD time delay of 1218 days of \citet{oguri2010} that is slightly smaller than the lower limit reported by \citet{fohl2008a}.\\

\begin{figure}
	\centering
	\includegraphics[width=8.5cm]{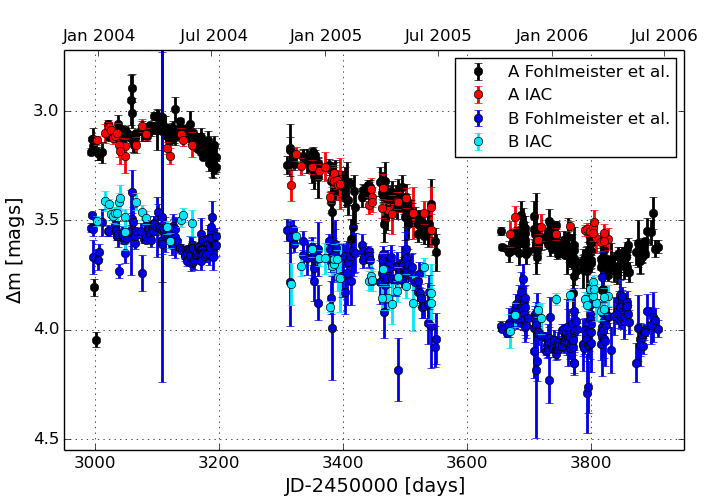}
  \caption{Light curves of the images A and B of the quasar SDSS J1004+4112 from December 2003 to June 2006. Horizontal axes show the Julian (bottom) and Gregorian (upper) dates. The IAC data is shifted to match the photometry by \citet{fohl2008a} to enable a direct comparison.\\
  }
	\label{Lens1004R_others}
\end{figure}

Fig. \ref{Lens1004R_others} shows a comparison of the IAC data and the data used by Fohlmeister et al. (2008) for the quasar images A and B. The light curves have, in general, a good agreement with those of \citet{fohl2008a} after applying global shifts to match both photometries in the overlapping regions. The exception is a region of image B at JD 3750 where the data of Fohlmeister et al. seem to be slightly below the expected values if one compare with the light curve for image A.\\

\section{Intrinsic Variability and Microlensing}
\begin{figure*}
\centering
\includegraphics[width=15cm]{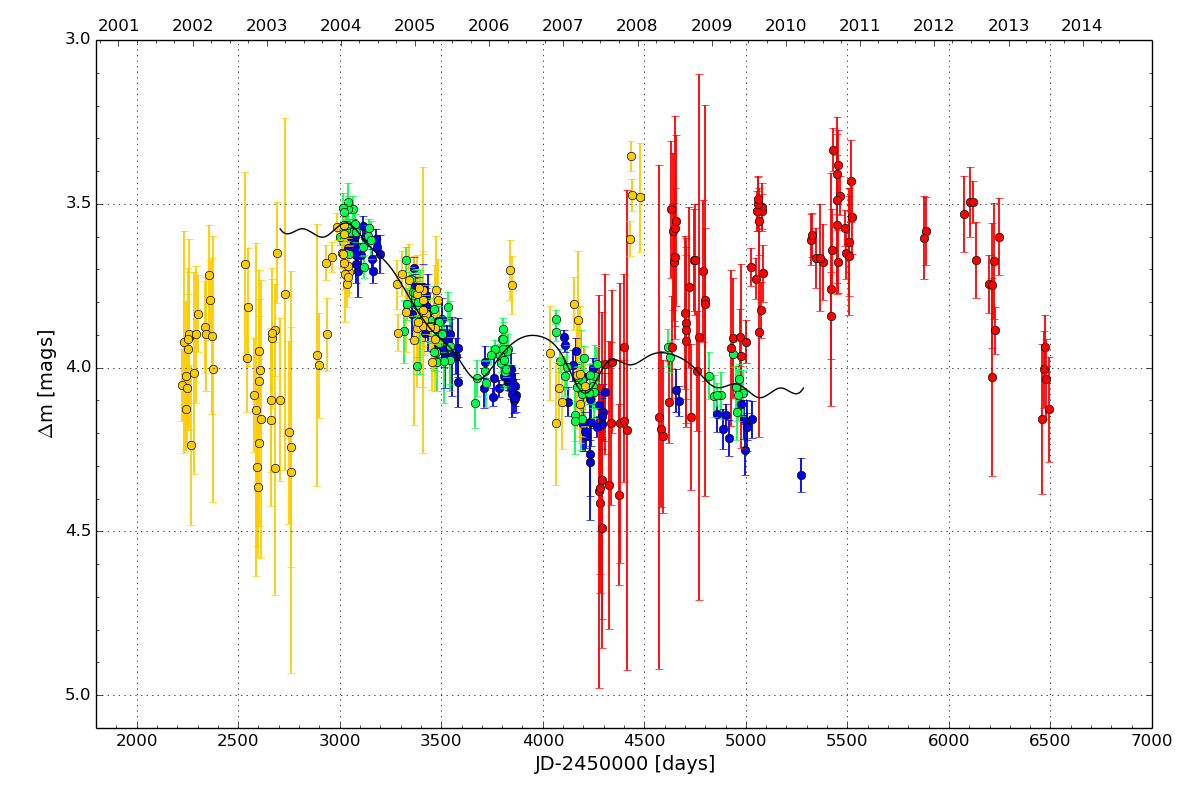}
\caption{Image A, B, C and D light curves of SDSS J1004+4112 in their overlap region after shifting by the respective time delays (and magnitude differences). A polynomial (black solid line) is fitted to the B light curve. We smooth the data with a square filter of $\pm$ 5 days to reduce noise.\\
}
\label{spline}
\end{figure*}

\begin{figure}
\includegraphics[width=8.5cm]{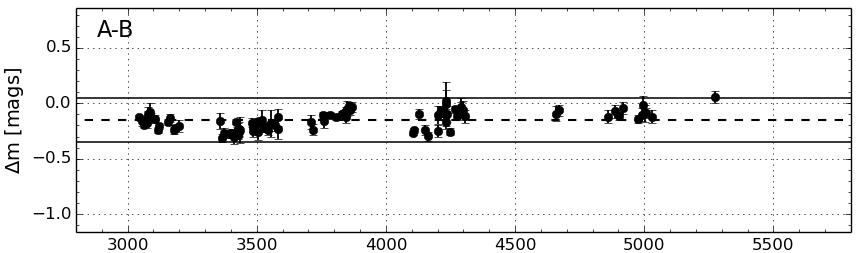}
\includegraphics[width=8.5cm]{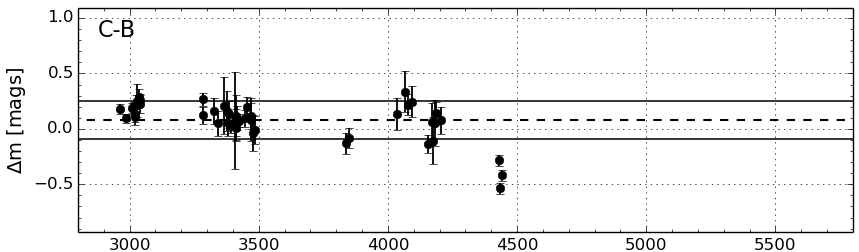}
\includegraphics[width=8.5cm]{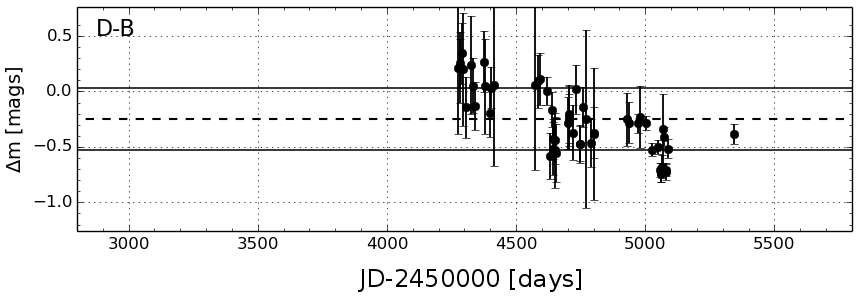}
\caption{Differential microlensing variability of the light curves A, C and D compared to a polynomial fit to the light curve B. The dashed horizontal lines show the mean value of the residuals. The residual magnitudes clearly show that microlensing is present in the light curves C and D.\\
}
\label{residuals}
\end{figure}

In Fig. \ref{Lens1004R_lightcurves} one can not see strong differences between the light curves A and B that could be directly related to microlensing. Thus, we can estimate the amplitude of the intrinsic variability of the quasar performing a single polynomial fitting to the B light curve. Despite the similarity between the A and B light curves, small microlensing could be still present due to a large source size or a location of both images in regions with low microlensing magnification, even strong microlensing magnification but with different sign in A and B. Consequently, although we use the polynomial fitting to make a tentative estimate for the intrinsic variability of the source, some contribution from microlensing variability cannot be completely discarded. Hence, we should compare with the A-B, C-B and D-B modeled magnification pattern/histograms for all the images. We obtain a source variability of $\sim$ 0.7 mag for all the data and of 0.5 mag for the first four seasons whereas \citet{fohl2008a} found an intrinsic variability of $\sim$ 0.7 mag. Fig. \ref{spline} shows the simulated quasar variability (black solid line) with the A, B, C and D light curves overimposed. In the three panels in Fig. \ref{residuals} the residuals of the A, C and D light curves are shown. We have calculated the residuals from: $\Delta m_X = m_X - m_{Bfit} - (m_X-m_B)_{midIR}$ where X = A, C, D. The midIR data have been taken from \citet{Ross2009}. The midIR emission is supposed to arise from a large enough region as to be not affected by microlensing and, hence, we use the $(m_x-m_B)_{midIR}$ offsets to set the microlensing baseline (see \citealt{Mediavilla2009}). In Fig. \ref{residuals} one can see that the residuals of the A light curve are relatively constant. Only in the fourth season, variability induced by microlensing may occur with amplitudes of order 0.15 mag. As opposed to this, microlensing variability is clearly visible in the residuals of the C and D light curves with amplitudes of the order of $\sim$ 0.5 mag and $\sim$ 0.7 mag. From these curves that represent the differential (with respect to B, the less prone to microlensing image) microlensing of the A, C and D images we have obtained the microlensing variability histograms (Fig. \ref{histograms}), i.e., the frequencies in which each microlensing amplitude appears in the microlensing variability light curves. We adopt a bin size of 0.2 mag.

\subsection{Microlensing Estimate of the Quasar Accretion Disk Size}
Taking into account that microlensing is sensitive to the size of the source (\citealt{Morgan2010}, see also the review by \citealt{wam2006}) we will use our determinations of microlensing magnification amplitude to estimate the size of the accretion disk in the SDSS J1004+4112 lensed quasar.\\ 

We obtained microlensing magnification maps for each image using the Inverse Polygon Mapping method described by \citet{Mediavilla2006,Mediavilla2011}. We used the following parameters obtained from a singular isothermal sphere plus external shear (SIS+$\gamma_e$) model fitted to the images coordinates: convergence $\kappa=0.48$ and shear $\gamma=0.57$ for image A, $\kappa=0.47$ and $\gamma=0.39$ for B, $\kappa=0.38$ and $\gamma=0.33$ for C and $\kappa=0.71$ and $\gamma=0.83$ for image D. We used a surface mass density in stars $\kappa_{*}$ of 10$\%$ \citep{Mediavilla2009} and generated 2000$x$2000 pixel$^2$ magnification maps with a size of 24$x$24 Einstein radii$^2$. The value of the Einstein radius for this system is $2.35 \times 10^{16}  \sqrt{M/0.3 M_\odot}$ cm = $9.1 \sqrt{M/0.3 M_\odot}$ light-days at the lens plane. The ratio of the magnification in a pixel to the average magnification of the map gives the microlensing magnification at the pixel and histograms of normalized to the mean maps deliver the relative frequency of microlensing magnification amplitude for a pixel-size source.\\

\begin{figure*}
\centering
\includegraphics[width=19cm]{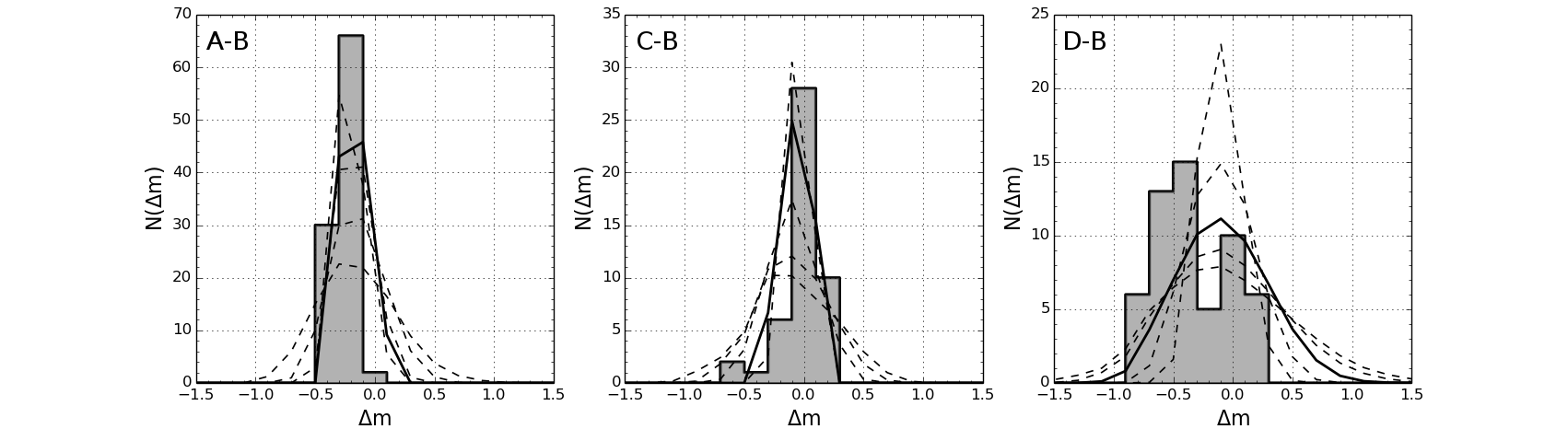}
\caption{Microlensing frequency distributions obtained from the observed light curves (histograms) and the simulated microlensing magnification maps. The polygonal lines present model histograms for different values of $r_s$ (6.3, 9.7, 15.0, 23.1 and 54.8 light-days for A-B, 4.1, 6.3, 9.7, 35.6 and 54.8 light-days for C-B, and 2.7, 4.1, 6.3, 9.7 and 15.0 light-days for D-B). The thick lines indicate the best fit of the model to the observed data.\\
}
\label{histograms}
\end{figure*}

To model the structure of the unresolved quasar source we considered a circular Gaussian intensity profile of size $r_s$, $I(R) \propto exp(-R^2/2 r_s ^2)$. It is generally accepted that the specific shape of the radial profile is not important for microlensing flux variability studies because the results are essentially controlled by the half-light radius rather than by the detailed profile \citep{Mortonson2005}. The characteristic size $r_s$ is related to the half-light radius by $R_{1/2} = 1.18 r_s$. We convolve the magnification maps with Gaussians of 14 different sizes over a logarithmic grid which spans approximately from $r_s$ $\sim$ 0.2 to 55 light-days for a mean stellar mass $<M>\ = 0.3 M_\odot$. The source sizes can be scaled to a different mean stellar mass, $\langle M \rangle$, using $r_s \propto \sqrt{M}$. After convolution we normalized each magnification map by its mean value. The histograms of the normalized map represent the histograms of the expected microlensing variability. Thus, we obtain 14 different microlensing histograms corresponding to different source sizes for each of the images A, C and D. Finally, convolving the histograms of A, C and D with the histogram of B we built the microlensing difference histograms A-B, C-B and D-B for different values of $r_s$ to be compared with the experimental histograms obtained from the observed light curves (see Fig. \ref{histograms}). The thick solid line in each panel indicates the best fit of the model to the observed frequency distributions of microlensing magnifications.\\

In order to study the likelihood of the different $r_s$ we compare the microlensing histograms inferred from the model for different values of $r_s$ with the histograms of the data using two different statistics: 
\begin{enumerate}
\item[(i)] histogram product, defined as 
\begin{equation}
P_X(r_s) = \sum_{i=1}^{n_{bin}} h_{X-B}^{i} \hat{h}_{X-B}^{i}(r_s),
\end{equation}
where $h_{X-B}^{i}$ and $\hat{h}_{X-B}^{i}(r_s)$ are the observed and modeled histograms and $n_{bin}$ is the number of bins
\item[(ii)] Pearson's $\chi^2$ test,
\begin{equation}
\chi^2 = \sum_{i=1}^{n_{bin}} \frac{(h_{X-B}^{i}-\hat{h}_{X-B}^{i}(r_s))^2}{{h_{X-B}^{i}}} 
\label{ns}
\end{equation}
with 
\begin{equation}
P_X(r_s) \propto e^{-\frac{1}{2} \chi^2}
\end{equation}
\end{enumerate}
After multiplying the probability distributions corresponding to A, C and D we finally obtain the PDF of the source size,
\begin{equation}
P(r_s) = P_A(r_s) \cdot P_C(r_s) \cdot P_D(r_s).
\label{prs}
\end{equation}

In the case of the histogram product method we have convolved the histograms with normal distributions with these dispersions not finding any significant difference.\\ 

To account for the rms errors in the data in the case of the $\chi^2$-method we applied a Monte Carlo method, producing random realizations of the histograms of the data using normal distributions of mean the value of the observed histogram and with the following dispersions: $\sigma_A=0.08$, $\sigma_C=0.14$ and $\sigma_D=0.22$ estimated from the data (see Fig. \ref{residuals}). In the case of the histogram product-method we have convolved the histograms with normal distributions with these dispersions not finding any significant difference.\\

\begin{figure}
\centering
\includegraphics[width=8.2cm]{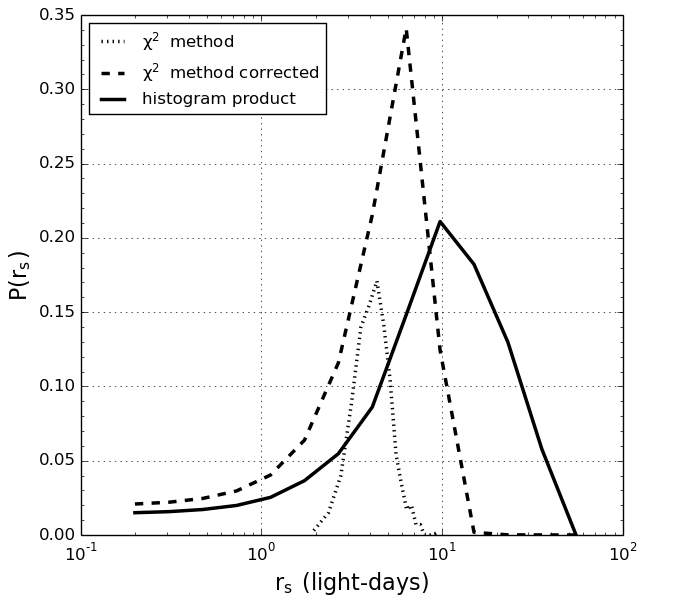}
\caption{Probability distributions of the source size $r_s$ for the product-method (solid line), $\chi^2$-method (dotted line) and $\chi^2$-method corrected with enlarged errors (dashed line).}
\label{disksize}
\end{figure}

Fig. \ref{disksize} shows the resulting normalized probability distribution for both methods. For the product we obtain a disk size of $<r_s>\ = 7.4^{+15.7}_{-4.7} \sqrt{M/0.3M_{\odot}}$. The expected value using $\chi^2$ is $<r_s>\ = 4.4^{+1.0}_{-1.0} \sqrt{M/0.3M_{\odot}}$ light-days (for 68$\%$ confidence estimates). The $\chi^2$-method seems to constrain better the size although the errors in the histograms may have been underestimated. This is supported by the rather high values of $\chi^2$. To reach values of $\chi^2$ close to 1 we need to multiply the typical deviation in Eq. \ref{ns}, $\sqrt{h_{X-B}^{i}}$ , by a factor 3. From the corresponding PDF (see Fig. \ref{disksize}) we obtain a similar value $<r_s>\ = 3.6^{+2.7}_{-1.9} \sqrt{M/0.3M_{\odot}}$ but with uncertainties more compatible with the results obtained using the product method.\\

Our result for each method expressed in terms of the half-light radius at $\lambda_{rest} = 2407\ \AA $ ($R_{1/2}=8.7^{+18.5}_{-5.5} \sqrt{M/0.3 M_{\odot}}$ and $R_{1/2} = 4.2^{+3.2}_{-2.2} \sqrt{M/0.3 M_{\odot}}$ light-days) is significantly greater than the value obtained by \citet{fohl2008a} fitting the A and B light curves ($R_{1/2}=2.44\ R_\lambda=0.6^{+0.5}_{-0.3}$ light days), but in good agreement with the value inferred by these authors from thin disk theory using an experimental estimate of the black hole mass \citep{fohl2008a}. Our estimate for the size is also in good agreement with the results by \citet{Motta2012} and \citet{Jimenez2014} for this system, and with the average determinations obtained for a sample of lensed quasars by \citet{Jimenez2012,Jimenez2014,Jimenez2015a,Jimenez2015b} when a fraction of mass in stars of 10\%  is considered.\\

As we have used an estimated (not measured) time delay for image D, and this image presents the largest microlensing signal in this system, it is important to check that this assumption is not biasing our results. In order to check the dependence of the results with the adopted time delay for image D, we have considered $\pm$ 100 days shifts of the time delays not finding any significant difference  in the size calculations. On the other hand, we have also repeated the computations not considering image D finding slightly larger but comparable results for the size: $<r_s>\ = 9.3^{+13.8}_{-5.2} \sqrt{M/0.3M_{\odot}}$ for the product-method and $<r_s>\ = 5.4^{+4.3}_{-2.7} \sqrt{M/0.3M_{\odot}}$ for $\chi^2$.\\

We have also checked the impact of interchanging the two images less affected by microlensing, A and B, performing the polynomial fitting to the A light curve and considering the pairs, B-A, C-A and D-A in the calculations. The results are almost identical for the $\chi^2$ based method, $<r_s>\ = 4.6^{+5.2}_{-1.9} \sqrt{M/0.3M_{\odot}}$, and compatible to within errors for the product method, $<r_s>\ = 4.6^{+5.1}_{-2.9} \sqrt{M/0.3M_{\odot}}$.

\section{Summary and Conclusions}
We have presented eight seasons of monitoring data for the four brightest images of the five image gravitational lens system SDSS J1004+4112 which significantly extend the
time coverage of previous works. Taken as reference image B that is less affected by microlensing and using the experimental time delays inferred by \citet{fohl2008a} for components A and C and the theoretical time delay modeled by \citet{oguri2010} for image D, we have removed the intrinsic variability from the light curves in the overlapping region. Using the mid-IR flux ratios between images determined by \citet{Ross2009} as a baseline for no microlensing magnification, we have finally obtained the microlensing light curves, A-B, C-B, and D-B. We have detected microlensing variability up to 0.7 mag in images C and D. The light curve of A seems to be less affected by microlensing (changes of order 0.15 mag in the fourth season of our data) in good agreement with the results of \citet{fohl2008a}.\\

We have used the statistics of microlensing magnifications along the available seasons to infer probabilistic distributions for the size using two different methods. Using the product of the observed and modeled microlensing histograms we have obtained a half-light radius of $R_{1/2}=8.7^{+18.5}_{-5.5} \sqrt{M/0.3M_\odot}$ light-days. A consistent but more restrictive result of $R_{1/2} = 4.2^{+3.2}_{-2.2} \sqrt{M/0.3M_\odot}$ light-days is obtained using a $\chi^2$ criterion to compare the observed and modeled histograms. Both results are only marginally consistent with the previous estimate by \citet{fohl2008a} that obtained a smaller size, but in good agreement with the predictions of thin disk theory and with the measurements by \citet{Motta2012} and \citet{Jimenez2014}.

\acknowledgments
We thank the anonymous referee for a thorough revision and valuable suggestions. We are especially grateful to the Instituto de Astrofísica de Canarias' astronomers for their observations of the data appearing in this paper. The 0.82m IAC80 Telescope is operated on the island of Tenerife by the Instituto de Astrofísica de Canarias in the Spanish Observatorio del Teide. J.J.V. is supported by the Spanish Ministerio de Econom\'\i a y Competitividad and the Fondo Europeo de Desarrollo Regional (FEDER) through grant AYA2014-53506-P and by the Junta de Andaluc\'\i a through project FQM-108.E.M. and J.A.M. were supported by the Spanish MINECO with the grants AYA2013-47744-C3-1-P and AYA2013-47744-C3-3-P. J.A.M. was also supported by the Generalitat Valenciana with the project PROMETEOII/2014/060.\\

\bibliographystyle{apj} % style aa.bst
\bibliography{bib} % your references Yourfile.bib

\begin{thebibliography}{37}
\expandafter\ifx\csname natexlab\endcsname\relax\def\natexlab#1{#1}\fi

\bibitem[{{Blackburne} {et~al.}(2014){Blackburne}, {Kochanek}, {Chen}, {Dai},
  \& {Chartas}}]{Blackburne2014}
{Blackburne}, J.~A., {Kochanek}, C.~S., {Chen}, B., {Dai}, X., \& {Chartas}, G.
  2014, \apj, 789, 125

\bibitem[{{Blackburne} {et~al.}(2015){Blackburne}, {Kochanek}, {Chen}, {Dai},
  \& {Chartas}}]{Blackburne2015}
---. 2015, \apj, 798, 95

\bibitem[{{Blackburne} {et~al.}(2011){Blackburne}, {Pooley}, {Rappaport}, \&
  {Schechter}}]{Blackburne2011}
{Blackburne}, J.~A., {Pooley}, D., {Rappaport}, S., \& {Schechter}, P.~L. 2011,
  \apj, 729, 34

\bibitem[{{Chang} \& {Refsdal}(1979)}]{Chang1979}
{Chang}, K. \& {Refsdal}, S. 1979, \nat, 282, 561

\bibitem[{{Chang} \& {Refsdal}(1984)}]{Chang1984}
---. 1984, \aap, 132, 168

\bibitem[{{Fohlmeister}(2008)}]{fohl2008b}
{Fohlmeister}, J. 2008, in Manchester Microlensing Conference, 16

\bibitem[{{Fohlmeister} {et~al.}(2008){Fohlmeister}, {Kochanek}, {Falco},
  {Morgan}, \& {Wambsganss}}]{fohl2008a}
{Fohlmeister}, J., {Kochanek}, C.~S., {Falco}, E.~E., {Morgan}, C.~W., \&
  {Wambsganss}, J. 2008, \apj, 676, 761

\bibitem[{{Fohlmeister} {et~al.}(2007){Fohlmeister}, {Kochanek}, {Falco},
  {Wambsganss}, {Morgan}, {Morgan}, {Ofek}, {Maoz}, {Keeton}, {Barentine},
  {Dalton}, {Dembicky}, {Ketzeback}, {McMillan}, \& {Peters}}]{fohl2007}
{Fohlmeister}, J., {Kochanek}, C.~S., {Falco}, E.~E., {Wambsganss}, J.,
  {Morgan}, N., {Morgan}, C.~W., {Ofek}, E.~O., {Maoz}, D., {Keeton}, C.~R.,
  {Barentine}, J.~C., {Dalton}, G., {Dembicky}, J., {Ketzeback}, W.,
  {McMillan}, R., \& {Peters}, C.~S. 2007, \apj, 662, 62

\bibitem[{{Hainline} {et~al.}(2013){Hainline}, {Morgan}, {MacLeod}, {Landaal},
  {Kochanek}, {Harris}, {Tilleman}, {Goicoechea}, {Shalyapin}, \&
  {Falco}}]{Hainline2013}
{Hainline}, L.~J., {Morgan}, C.~W., {MacLeod}, C.~L., {Landaal}, Z.~D.,
  {Kochanek}, C.~S., {Harris}, H.~C., {Tilleman}, T., {Goicoechea}, L.~J.,
  {Shalyapin}, V.~N., \& {Falco}, E.~E. 2013, \apj, 774, 69

\bibitem[{{Inada} {et~al.}(2005){Inada}, {Oguri}, {Keeton}, {Eisenstein},
  {Castander}, {Chiu}, {Hall}, {Hennawi}, {Johnston}, {Pindor}, {Richards},
  {Rix}, {Schneider}, \& {Zheng}}]{inada2005}
{Inada}, N., {Oguri}, M., {Keeton}, C.~R., {Eisenstein}, D.~J., {Castander},
  F.~J., {Chiu}, K., {Hall}, P.~B., {Hennawi}, J.~F., {Johnston}, D.~E.,
  {Pindor}, B., {Richards}, G.~T., {Rix}, H.-W.~R., {Schneider}, D.~P., \&
  {Zheng}, W. 2005, \pasj, 57, L7

\bibitem[{{Inada} {et~al.}(2006){Inada}, {Oguri}, {Morokuma}, {Doi}, {Yasuda},
  {Becker}, {Richards}, {Kochanek}, {Kayo}, {Konishi}, {Utsunomiya}, {Shin},
  {Strauss}, {Sheldon}, {York}, {Hennawi}, {Schneider}, {Dai}, \&
  {Fukugita}}]{inada2006}
{Inada}, N., {Oguri}, M., {Morokuma}, T., {Doi}, M., {Yasuda}, N., {Becker},
  R.~H., {Richards}, G.~T., {Kochanek}, C.~S., {Kayo}, I., {Konishi}, K.,
  {Utsunomiya}, H., {Shin}, M.-S., {Strauss}, M.~A., {Sheldon}, E.~S., {York},
  D.~G., {Hennawi}, J.~F., {Schneider}, D.~P., {Dai}, X., \& {Fukugita}, M.
  2006, \apjl, 653, L97

\bibitem[{{Jim{\'e}nez-Vicente}
  {et~al.}(2015{\natexlab{a}}){Jim{\'e}nez-Vicente}, {Mediavilla}, {Kochanek},
  \& {Mu{\~n}oz}}]{Jimenez2015b}
{Jim{\'e}nez-Vicente}, J., {Mediavilla}, E., {Kochanek}, C.~S., \& {Mu{\~n}oz},
  J.~A. 2015{\natexlab{a}}, \apj, 799, 149

\bibitem[{{Jim{\'e}nez-Vicente}
  {et~al.}(2015{\natexlab{b}}){Jim{\'e}nez-Vicente}, {Mediavilla}, {Kochanek},
  \& {Mu{\~n}oz}}]{Jimenez2015a}
---. 2015{\natexlab{b}}, \apj, 806, 251

\bibitem[{{Jim{\'e}nez-Vicente} {et~al.}(2014){Jim{\'e}nez-Vicente},
  {Mediavilla}, {Kochanek}, {Mu{\~n}oz}, {Motta}, {Falco}, \&
  {Mosquera}}]{Jimenez2014}
{Jim{\'e}nez-Vicente}, J., {Mediavilla}, E., {Kochanek}, C.~S., {Mu{\~n}oz},
  J.~A., {Motta}, V., {Falco}, E., \& {Mosquera}, A.~M. 2014, \apj, 783, 47

\bibitem[{{Jim{\'e}nez-Vicente} {et~al.}(2012){Jim{\'e}nez-Vicente},
  {Mediavilla}, {Mu{\~n}oz}, \& {Kochanek}}]{Jimenez2012}
{Jim{\'e}nez-Vicente}, J., {Mediavilla}, E., {Mu{\~n}oz}, J.~A., \& {Kochanek},
  C.~S. 2012, \apj, 751, 106

\bibitem[{{Kochanek}(2004)}]{Kochanek2004}
{Kochanek}, C.~S. 2004, \apj, 605, 58

\bibitem[{{Landolt}(1992)}]{Landolt1992}
{Landolt}, A.~U. 1992, \aj, 104, 340

\bibitem[{{MacLeod} {et~al.}(2015){MacLeod}, {Morgan}, {Mosquera}, {Kochanek},
  {Tewes}, {Courbin}, {Meylan}, {Chen}, {Dai}, \& {Chartas}}]{MacLeod2015}
{MacLeod}, C.~L., {Morgan}, C.~W., {Mosquera}, A., {Kochanek}, C.~S., {Tewes},
  M., {Courbin}, F., {Meylan}, G., {Chen}, B., {Dai}, X., \& {Chartas}, G.
  2015, \apj, 806, 258

\bibitem[{{Mediavilla} {et~al.}(2015){Mediavilla}, {Jim{\'e}nez-vicente},
  {Mu{\~n}oz}, \& {Mediavilla}}]{Mediavilla2015}
{Mediavilla}, E., {Jim{\'e}nez-vicente}, J., {Mu{\~n}oz}, J.~A., \&
  {Mediavilla}, T. 2015, \apjl, 814, L26

\bibitem[{{Mediavilla} {et~al.}(2011){Mediavilla}, {Mediavilla}, {Mu{\~n}oz},
  {Ariza}, {Lopez}, {Gonzalez-Morcillo}, \& {Jimenez-Vicente}}]{Mediavilla2011}
{Mediavilla}, E., {Mediavilla}, T., {Mu{\~n}oz}, J.~A., {Ariza}, O., {Lopez},
  P., {Gonzalez-Morcillo}, C., \& {Jimenez-Vicente}, J. 2011, \apj, 741, 42

\bibitem[{{Mediavilla} {et~al.}(2009){Mediavilla}, {Mu{\~n}oz}, {Falco},
  {Motta}, {Guerras}, {Canovas}, {Jean}, {Oscoz}, \&
  {Mosquera}}]{Mediavilla2009}
{Mediavilla}, E., {Mu{\~n}oz}, J.~A., {Falco}, E., {Motta}, V., {Guerras}, E.,
  {Canovas}, H., {Jean}, C., {Oscoz}, A., \& {Mosquera}, A.~M. 2009, \apj, 706,
  1451

\bibitem[{{Mediavilla} {et~al.}(2006){Mediavilla}, {Mu{\~n}oz}, {Lopez},
  {Mediavilla}, {Abajas}, {Gonzalez-Morcillo}, \&
  {Gil-Merino}}]{Mediavilla2006}
{Mediavilla}, E., {Mu{\~n}oz}, J.~A., {Lopez}, P., {Mediavilla}, T., {Abajas},
  C., {Gonzalez-Morcillo}, C., \& {Gil-Merino}, R. 2006, \apj, 653, 942

\bibitem[{{Morgan} {et~al.}(2010){Morgan}, {Kochanek}, {Morgan}, \&
  {Falco}}]{Morgan2010}
{Morgan}, C.~W., {Kochanek}, C.~S., {Morgan}, N.~D., \& {Falco}, E.~E. 2010,
  \apj, 712, 1129

\bibitem[{{Mortonson} {et~al.}(2005){Mortonson}, {Schechter}, \&
  {Wambsganss}}]{Mortonson2005}
{Mortonson}, M.~J., {Schechter}, P.~L., \& {Wambsganss}, J. 2005, \apj, 628,
  594

\bibitem[{{Mosquera} \& {Kochanek}(2011)}]{Mosquera2011}
{Mosquera}, A.~M. \& {Kochanek}, C.~S. 2011, \apj, 738, 96

\bibitem[{{Mosquera} {et~al.}(2013){Mosquera}, {Kochanek}, {Chen}, {Dai},
  {Blackburne}, \& {Chartas}}]{Mosquera2013}
{Mosquera}, A.~M., {Kochanek}, C.~S., {Chen}, B., {Dai}, X., {Blackburne},
  J.~A., \& {Chartas}, G. 2013, \apj, 769, 53

\bibitem[{{Mosquera} {et~al.}(2009){Mosquera}, {Mu{\~n}oz}, \&
  {Mediavilla}}]{Mosquera2009}
{Mosquera}, A.~M., {Mu{\~n}oz}, J.~A., \& {Mediavilla}, E. 2009, \apj, 691,
  1292

\bibitem[{{Motta} {et~al.}(2012){Motta}, {Mediavilla}, {Falco}, \&
  {Mu{\~n}oz}}]{Motta2012}
{Motta}, V., {Mediavilla}, E., {Falco}, E., \& {Mu{\~n}oz}, J.~A. 2012, \apj,
  755, 82

\bibitem[{{Mu{\~{n}}oz} {et~al.}(2016){Mu{\~{n}}oz}, {Vives-Arias}, {Mosquera},
  {Jim{\'e}nez-Vicente}, {Kochanek}, \& {Mediavilla}}]{Munoz2015}
{Mu{\~{n}}oz}, J.~A., {Vives-Arias}, H., {Mosquera}, A.~M.,
  {Jim{\'e}nez-Vicente}, J., {Kochanek}, C.~S., \& {Mediavilla}, E. 2016, \apj,
  817, 155

\bibitem[{{Oguri}(2010)}]{oguri2010}
{Oguri}, M. 2010, \pasj, 62, 1017

\bibitem[{{Pooley} {et~al.}(2007){Pooley}, {Blackburne}, {Rappaport}, \&
  {Schechter}}]{Pool2007}
{Pooley}, D., {Blackburne}, J.~A., {Rappaport}, S., \& {Schechter}, P.~L. 2007,
  \apj, 661, 19

\bibitem[{{Refsdal}(1964)}]{Ref1964}
{Refsdal}, S. 1964, \mnras, 128, 307

\bibitem[{{Ross} {et~al.}(2009){Ross}, {Assef}, {Kochanek}, {Falco}, \&
  {Poindexter}}]{Ross2009}
{Ross}, N.~R., {Assef}, R.~J., {Kochanek}, C.~S., {Falco}, E., \& {Poindexter},
  S.~D. 2009, \apj, 702, 472

\bibitem[{{Serra-Ricart} {et~al.}(1999){Serra-Ricart}, {Oscoz},
  {Sanch{\'{\i}}s}, {Mediavilla}, {Goicoechea}, {Licandro}, {Alcalde}, \&
  {Gil-Merino}}]{Serra1999}
{Serra-Ricart}, M., {Oscoz}, A., {Sanch{\'{\i}}s}, T., {Mediavilla}, E.,
  {Goicoechea}, L.~J., {Licandro}, J., {Alcalde}, D., \& {Gil-Merino}, R. 1999,
  \apj, 526, 40

\bibitem[{{Sluse} {et~al.}(2011){Sluse}, {Schmidt}, {Courbin},
  {Hutsem{\'e}kers}, {Meylan}, {Eigenbrod}, {Anguita}, {Agol}, \&
  {Wambsganss}}]{Sluse2011}
{Sluse}, D., {Schmidt}, R., {Courbin}, F., {Hutsem{\'e}kers}, D., {Meylan}, G.,
  {Eigenbrod}, A., {Anguita}, T., {Agol}, E., \& {Wambsganss}, J. 2011, \aap,
  528, A100

\bibitem[{{Wambsganss}(2003)}]{wam2003}
{Wambsganss}, J. 2003, \nat, 426, 781

\bibitem[{{Wambsganss}(2006)}]{wam2006}
---. 2006, ArXiv Astrophysics e-prints

\end{thebibliography}
\end{document}